\documentclass[12pt,epsfig,multirow]{article}
\usepackage{graphicx}
\usepackage[usenames]{color}
\newcommand{\be}{\begin{eqnarray}}
\newcommand{\ee}{\end{eqnarray}}

\def\dis{\displaystyle}
\def\barr{\begin{array}}
\def\earr{\end{array}}

\begin{document}

\thispagestyle{empty}
\begin{flushright}

\end{flushright}
\bigskip

\begin{center}
{\Large \bf Study of isospin nonconservation in the framework of spectral distribution theory}

\vspace{.5in}

{\bf {Kamales Kar$^{\star, a}$ and Sukhendusekhar Sarkar$^{\dagger,b}$}}
\vskip .5cm
$^\star${\normalsize \it Ramakrishna Mission Vivekananda University,}
{\normalsize \it Belur Math, Howarah 711202, India}\\
\vspace{0.05in}
$^\dagger${\normalsize \it Indian Institute of Engineering Science and Technology, Shibpur, Howrah 711103, India}\\
\vskip 0.4cm

\vskip 1cm

{\bf ABSTRACT}
\end{center}
The observed isospin-symmetry breaking in light nuclei are caused not only by
the Coulomb interaction but by the isovector one- and two-body plus isotensor two-
body nuclear interactions as well. Spectral distribution theory which treats
nuclear spectroscopy and other structural properties in a statistical
framework was earlier applied mostly to
isospin conserving Hamiltonians. In this paper we extend that to include
the nuclear interactions non-scalar in isospin and work out examples in sd
shell to calculate the linear term in the isobaric mass-multiplet equation
originating from these non-isoscalar parts.

\vfill

\noindent $^a$ email: kamales.kar@gmail.com

\noindent $^b$ email: ss@physics.becs.ac.in ; sukhendusekhar.sarkar@gmail.com

\section{Introduction}                         \label{sec:introduction}
Presently, many studies based on the no-core shell models or valence shell models, utilise 
charge dependent interactions,working in the proton-neutron formalism. The resulting wave functions ,
 in these calculations after diagonalisation, possess a slight isospin admixture. However, in
 conventional nuclear structure physics, isospin is often considered to be a good
symmetry, at least as a starting point, with the isospin operator {\bf T} commuting with the nuclear one and
two-body Hamiltonian ($H_{nuc}(1)$ and $H_{nuc}(2)$) so that the many-nucleon
energy eigenstates are labeled by specific isospin quantum numbers $T$ and they
have a degeneracy of $(2T+1)$. The states of the multiplet i.e.
$T_z= -T, -T+1,...T-1, T$ belong to different nuclei and are often called
the Isobaric Analog States (IAS). This follows from the postulate of
Charge Independence of the interaction meaning that the proton-proton,
proton-neutron and neutron-neutron interactions are identical. Extensive
experimental evidence gathered by now confirms that isospin symmetry is
broken by not only the electromagnetic Coulomb interaction but also by a
small amount of charge-dependent nuclear interaction \cite{Lam-12}. The difference of energies
of the ground states (or specific low-lying states) of two mirror nuclei i.e.
nuclei with neutron number (N) and proton number (Z) interchanged, is dominantly due to
the isospin nonconservation and isoscalar part of the coulomb interaction.
A number of examples, particularly of light nuclei are available with
reasonably accurate measurements of the Mirror Energy Difference (MED)
\cite{Jenkins-05} given by $\Delta E=E(J,T_z=-T) -E(J,T_z=+T)$ with T=1/2, 1....
Actually Wigner \cite{Wigner-57} had postulated an isobaric mass multiplet
equation (IMME) to take care of the splitting, in general, going upto
the quadratic term as
\be
M(\alpha,T,T_z)=a(\alpha,T)+ b(\alpha,T) T_z+ c(\alpha,T) T_z^2
\ee

for multiplets for given $\alpha$ and $T$. Here $\alpha$ represents all other relevant quantum numbers including the nuclear mass $A$.
 The coefficients $a$, $b$ and $c$
can be calculalted by theory or extracted from experimental results. On
 the theoretical side one has constructed one and two-body Hamiltonians
 no longer isoscalar but with isovector and isotensor parts as well
and performed shell model \cite{Ormand-89} and other calculations.
The parameters of the one plus two-body interaction used in the shell model
calculations were then best fitted to reproduce the experimental data
\cite{Lam-12}.

In contrast to microscopic structure models like the nuclear shell model a
theory to describe the statistically averaged energy spectra and transition
strengths of nuclear excitations and decays have been developed over the years
based on the results of random matrix ensembles embedded in many-nucleon spaces \cite{French-1966,French-67,FR-71,
CFT-71,kothq-10}.
This theory \cite{Brody-81,French-82,Kota-89}
avoids the diagonalisation of large Hamiltonian matrices and use
the low moments of the Hamiltonian by evaluating the traces of low powers
of the Hamiltonian in many-nucleon spaces. The basic
result of this theory is the fact that the averaged density of energy
eigenstates goes fast towards a Gaussian with the increase in the number of
valence nucleons in large dimensional shell model spaces. This result is seen to
be true even for each configuration when one considers configuration
partitioning as long as the configuration dimension is large. Here configuration partitioning 
means distributing in various ways `m' valence particles over the `1' valence orbits ($j_i$) coupled to final isospin T, that is,
${(\bf m},T)$ = $({m_1,m_2,....,m_l})^{T}$ =
$[{j_1}^{ m_1}
 \otimes {j_2}^{m_2}$ $......\otimes {j_l}^{ m_l}]^{T}$ with $ m =\sum\limits_{i=1}^{l}{m_i}$. 
Though in
principle this theory is not meant for considering individual states,
even then one can invert the problem and locate low-lying discrete states in
energy like the ground state and compare the values with experimental binding
energies. This has yielded reasonable success for light nuclei not only for the
stable ones \cite{Sarkar-87}, \cite{Kar-97} ,\cite{Choubey-98} but for the neutron-rich nuclei as well
\cite{Kar-13}. First studies of isospin nonconserving interactions and their effect on nuclei using 
spectral distribution theory were carried forward by Hecht and Draayer\cite{hd-74}, following earlier
discussion by French[18] 
 But then almost all applications of spectral
 distribution theory have been done considering isospin conserving
interaction only. In this work we plan to extend this to include
isovector and isotensor Hamiltonians and describe how the relevant traces
are first evaluated in spaces with many neutrons and protons and then
through subtraction of these traces one projects out the ones in states with
fixed isspin. This is then used to estimate the coefficient $b$ in IMME.
In section 2 some basic features of the spectral distribution theory
with respect to the calculation of low-lying spectra are described.
Section 3 involves spectral distribution investigations including the
non-isoscalar Hamiltonians and presents some results for $sd$ shell nuclei.
Section 4 gives some concluding remarks.

\section{Methods of Spectral Distribution Theory}    \label{sec:be}

The Gaussian nature of the density of states can be demonstrated for the
non-interacting case, i.e. for a  one-body Hamiltonian, by using the
Central Limit Theorem
(CLT) neglecting the Pauli blocking effect. For the interacting
case i.e. with (1+2)-body Hamiltonian, one needs to carry out an averaging
over the Embedded Gaussian Orthogonal Ensemble (EGOE) \cite{Mon-73}
\cite{Gomez-11} to get the Gaussian result.
With `m' valence particles distributed over `N' single particle states
giving the shell-model space dimension $d(m)$ (equal to
 $^{N}C_{m}$), the
normalised density of states $\rho_{m}(E)$ is then a function only of the
centroid $\epsilon(m)$ and width $\sigma(m)$ of the Gaussian.
But for using the result for real nuclei one needs to partition the space
into subspaces with fixed isospin and for better accuracy, do a
configuration partitioning, by distributing the `m' particles in `l'
shell model orbits, giving rise to the normalised configuration-isospin
densities, $\rho_{{\bf m},T}(E)$ and dimensions $d({\bf m},T)$, where, 
${\bf m}$ = (${m_1,m_2,....,m_l}$). The dimension of
the scalar-isospin space with m valence particles and isospin T and with a specific value for $T_z$, is given by 

\be
d(m,T)= \frac{2(2T+1)}{N+2}{\frac{1}{2}N+1 \choose \frac{1}{2}m-T}{\frac{1}{2}N+1 \choose \frac{1}{2}m+T+1}
\ee
 where N is the total number of single particle states. For the expression for configuration-isospin space
dimension we refer to Kar[21].


The Gaussian form of the eigenvalue
density is seen to be true in the $({\bf m},T)$ subspaces too as long as
dimension of each subspace is large enough. The intensities ($I_{\bf m,T} (E)$) in the
subspaces then just add up to give the total intensity. 

\be
I_{m, T}(E)= \dis \sum_{\bf m} I_{{\bf m},T} (E)= \dis \sum_{\bf m} d({\bf m},T) \rho_{{\bf m},T} (E)
\ee

The asymptotic Gaussian result is observed for almost all realistic
Hamiltonians and the calculated higher cumulants are found to be small. For the second moment
in the space with fixed values of valence particles `m' and isospin `T', one writes for width squared
$\sigma^2 (m,T) =$  $<H^2>^{m,T} - [<H>^{m,T}]^2$
where the first term on the RHS is the square  of the norm of H and the second is the square of the centroid evaluated in the (m,T) space. The
 widths in configuration spaces have similar expressions.

The ground state energy $\bar{E_{g}}$ of a nucleus with the number of
valence nucleons $m$ and isospin $T$ is given by the Ratcliff prescription \cite{Ratcliff-71}

\be
\dis \sum_{\bf{m}} \int_{-\infty}^{\bar{E_{g}}} I_{{\bf m},T}(E) dE = d_{0}/2
\ee

where $d_{0}$ is the degeneracy of the ground state. Thus the energy where the
integrated area below the level density from the low energy side reaches half
the ground state degeneracy, is taken to be the energy of the ground state.

The total Hamiltonian under consideration here is given by

\be
H= \dis \sum_{\nu} H^{(\nu)} = H^{(0)} +H^{(1)} +H^{(2)}
\ee

where $H^{(0)}$, $H^{(1)}$ and $H^{(2)}$ are the isoscalar, isovector and
isotensor (i.e. tensor of rank two in isospin) parts.
For handling this Hamiltonian one first writes all the three parts
in proton-neutron(p,n) formalism.
The proton and neutron single particle matrix elements are given by
\cite{Ormand-89} $\epsilon^{p}_{r}=\epsilon^{0}_{r}+\epsilon^{1}_{r}/2$
 and $\epsilon^{n}_{r}=\epsilon^{0}_{r}-\epsilon^{1}_{r}/2$ where
$\epsilon^{0}_{r}$ and $\epsilon^{1}_{r}$ are the isoscalar and isovector
single particle matrix elements for orbit `r'. The values of the
isovector and isotensor single particle energies and two-body matrix elements
were obtained by Ormand and Brown \cite{Ormand-89}
for the  $0p$, $pds$, $1s-0d$, $df$ and $0f-1p$ shells by a least squared fit
to the observed `b' and `c' coefficients of the IMME. We use the same values
here though we limit ourselves to the study of the coefficient `b' only.
The relation between the two-body matrix elements in the (p,n) formalism
and the fixed isospin formalism is well known \cite{French-66} and so we
mention it here briefly. The
proton-proton, neutron-neutron along with the
neutron-proton or proton-neutron `T=1' part matrix elements are \cite{Ormand-89}

\be
V^{pp}_{rstu:J}=V^{(0)}_{rstu;J,T=1} +V^{(1)}_{rstu;J,T=1}/2 +V^{(2)}_{rstu;J,T=1}/6
\ee

\be
V^{nn}_{rstu;J}=V^{(0)}_{rstu;J,T=1} -V^{(1)}_{rstu;J,T=1}/2 +V^{(2)}_{rstu;J,T=1}/6
\ee

\be
V^{pn}_{rstu;J}(T=1)=V^{(0)}_{rstu;J,T=1}-V^{(2)}_{rstu;J,T=1}/3
\ee

where $V^{(\nu)}_{rstu;J,T}$ are the two-body matrix elements in the isospin
formalism. The $\nu$ =0,1,2 stand for the isoscalar, isovector and rank 2 isotensor
matrix elements respectively. Then the total neutron-proton or proton-neutron
two-body matrix  elements become \cite{Ormand-89}

\be
V^{pn}_{rstu;J}=[(1+\delta_{rs})(1+\delta_{tu})/2)]^{1/2} (V^{pn}_{rstu;J}(T=1)+{V^{(0)}}_{rstu;J,T=0})
\ee
where ${V^{pn}}_{rstu} (T=1)$ is the proton-neutron matrix elemen given in equation (8).

In the spectral distribution method
the traces of one operator or a product of operators in second quantized form
are obtained by the contraction
of the creation and destruction operators and then summing over all states.
Alternatively one can write expressions for the trace equivalent operators
in a closed form using the scalars of the relevant group involved
\cite{French-66}, \cite{Kar-83}.
For example, the trace equivalent(TE) Hamiltonian that reproduces centroids in
the  `pn' subspaces is given in terms of the scalars of
$\Sigma_{r} [U(N_{r}/2)+ U(N_{r}/2)]$ subgroup by

\vskip -0.2cm
\be
\begin{array}{rcl}
H(TE;pn)& = & \dis \sum_{r} \epsilon^{p}_{r} n^{p
}_{r} +\dis \sum_r \epsilon^{n}_r
n^{n}_{r} +\dis \sum_{r} W^{pp}_{rr} n^{p}_{r} (n^{p}_{r}-1)/2
+\dis \sum_{r< s} W^{pp}_{rs} n^{p}_{r} n^{p}_{s}   \\
& + & \dis \sum_{r} W^{nn}_{rr} n^{n}_{r} (n^{n}_{r}-1)/2
+\dis \sum_{r< s} W^{nn}_{rs} n^{n}_{r} n^{n}_{s}
+\dis \sum_{r,s} W^{pn}_{rs} n^{p}_{r} n^{n}_{s}
\end{array}
\ee

where $n^{p}_{r}$ and $n^{n}_{r}$ are the proton and neutron number operators
in orbit `r',
$W^{pp}_{rr}$ and $W^{nn}_{rr}$ are the averaged two-body proton-proton
and neutron-neutron matrix elements in orbit `r' and $W^{pp}_{rs}$,
$W^{nn}_{rs}$ and $W^{pn}_{rs}$ the averaged two-body proton-proton,
neutron-neutron,  proton-neutron two-body matrix elements with one particle
in orbit `r' and the other in orbit `s'.

For calculating the widths, we use the relation between the square of the norm of the traceless part of the 
total Hamiltonian (\~{H}) and the products of the traceless parts of the isoscalar, isovector and isotensor Hailtonians, given by
\be
<\tilde{H}^{2}>= <\tilde{H}^{(0)}\tilde{H}^{(0)}>+ 2<\tilde{H}^{(0)}\tilde{H}^{(1)}>+ 2<\tilde{H}^{(0)}\tilde{H}^{(2)}>
\ee

where $<...>$ denote averages in the relevant spaces and we neglect the terms
quadratic involving $H^{(1)}$ or $H^{(2)}$ as the terms
non-scalar in isospin are small. Once the centroids and
variances in all the proton-neutron configuration spaces ($(\tilde{{\bf m}}_p$, $\tilde{{\bf m}}_n))$
 are evaluated for
a fixed number of particles, corresponding averages in fixed isospin
configuration spaces are then projected out by a procedure involving the (p,n)
 traces including the relevant Clebsch-Gordan coefficients for nonisoscalar
Hamiltonians. Briefly the method is as follows \cite{Mugambi-70},
\cite{Sarkar-92}.

If we consider a space with $m_{p}$ protons and $m_{n}$ neutrons then the
eigenvalue of $T_{z}$ is $T_{0}=(m_{n}-m_{p})/2$. Then the isospin quantum
number has values  $T_{0}$, $T_{0}+1$,...,$(m_{n}+m_{p})/2$. The trace of
operator $O$  in the (p,n) spaces is given as a sum over the
traces of the reduced matrix elements of $O^{\lambda_{T}}$ in the isospin
spaces with different $T$ values

\be
<<O>>^{m_n=m-k,m_p=k}=\dis \sum_{i=0}^{k} (m-2i+1)^{-1/2} (C^{m/2-i~\lambda_{T}~m/2-i}_{m/2-k~0~m/2-k}) <<O^{\lambda_{T}}>>^{m,T=m/2-i}
\ee

where $<<...>>^{m_{n},m_{p}}$ and $<<...>>^{m,T}$ are the traces in
the (p,n) spaces and isospin spaces respectively. 
Putting k=0,1,...m one gets a set of (m+1) equations. By inverting them one can write the  traces of the reduced matrix elements
in (m,T) spaces in terms of the traces in the (p,n) spaces.

\section{Applications to some nuclei in the $1s-0d$ shell} \label{section nct}

The spectral distribution results are obtained in configuration spaces
with fixed number of valence particles and isospin.
For that calculations are first carried out in proton-neutron configuration
spaces where instead of the shell model space of `l' orbits one considers
`2l' orbits with the first `l' as proton orbits and the next `l'
orbits as neutron orbits. The single particle energies(spe) similarly
have `2l' values the first `l' being proton single particle energies
and the next `l' as neutron single particle energies. When one has only
isoscalar interaction the proton  and neutron spe-s for a specific orbit are the
same but once one includes T-nonconserving interactions they are no longer the
same as can be seen from their expressions given in the previous section.
In the $1s-0d$ shell there are 6 orbits in this pn-formalism
with 6 distinct spe-s. The two-body matrix elements in the pn-formalism
are given,  instead of the
$V^{JT}_{rstu}$ where (JT) stand for angular momentum and isospin
and `r',`s',`t' and `u' for the orbits, in the form of
$V^{pp}_{rstu;J}$, $V^{nn}_{rstu;J}$ and $V^{pn}_{rstu;J}$  given
by equations (6-9).
For the isoscalar part of the interaction in $1s-0d$ shell we use the highly
successful Wildenthal's mass dependent interaction.
For the nonisoscalar one and two-body parts we use the values of Ormand and
Brown \cite{Ormand-89} as already mentioned because the shell model results done
 with them show good agreement with experiment.
The traces for the two lowest moments are first calculated in spaces with fixed proton and neutron numbers
and then the traces in the fixed isospin spaces are evaluated applying equation (12).

We consider three examples in  the $1s-0d$ shell nuclei and calculate the ground
state energies by spectral distributions with the total Hamiltonian
as well as with only the isoscalar and isovector parts. From the contribution of the isovector part one can
estimate the parameter `b' of IMME coming from the nuclear interactions. We add to that
the Coulomb contribution to `b' taking  values from
figures 9 and 10 of \cite{Lam-12} for all the 3 cases.

Table 1 and Table 2 give
the values for the centroids and widths for the isoscalar as well as for the total
Hamiltonian for nuclei with  4 and 5 valence partcles in $1s-0d$ shell
with isospin 1 and 3/2 respectively. They show all possible fixed (m,T) configurations and the notation
($m_1,m_2,m_3$) stands for $m_1$ particles in $0d_{5/2}$, $m_2$ particles in $0d_{3/2}$, and $m_3$ particles in $1s_{1/2}$.
All the results show a very interesting aspect: once one includes the
isospin violating parts in the interaction the centroid of each
configuration moves down by a few MeV whereas the width remains essentially
the same. Thus the whole configuration energy state
density just shifts in energy due to the isospin violating terms. One also realises that this change increases with the T value with the total number of valence particles remaining the same.
 This is mainly
because the nonzero isovector single particle energies contribute to the
centroids but has very little contribution in spreading the states in the spaces
 around the centroids. One also realises that this change increases with the T value with the total number of valence particles remaining the same. Table 3 shows
our results for the ground state energies (GSE) by spectral distributions with
and without the isospin violating parts and the value of the parameter `b'
calculated for the nuclear interactions. The GSE values are with respect to
$^{16}O$ as the closed core. The Coulomb contributions are then
added to these values. The results agree reasonably well with the observed
values which are also included in Table 3. The table gives the ground state energies for three nuclei for the cases of 
(i) only isoscalar Hamiltonian, (ii) isoscalar and isovector Hamiltonians (iii) the total Hamiltonian including isoscalar, isovector and isotensosor parts.
The difference of energies for the first two cases is equal to  bT ( where $T_z$ has the value T in the ground state). So this energy difference divided by T gives the value of the coefficient b and given in column 5 of table 3.
 Keeping in mind that the spectral
distribution theory is a statistical framework for the global properties of the
nuclei the agreement is satisfactory close to the observed value as well as those obtained in \cite{Lam-12}, \cite{ Ormand-89}.
We note here that Ormand \cite { Ormand-97} obtained in  his study a formula
for 'b' with a good global fit to data and that gives values for 'b' as 4.28, 4.46 and 4.63 for
the three cases with A=20, 21 and 22, respectively and this is again close to the values obtained in the SDT calculations.
 We also note that in earlier works
to predict the ground state energies more accurately one considered corrections
including low-lying excited states in the spectra as well as by including
small non-zero skewness and excess for the averaged configuration density
distributions \cite{Sarkar-87}, \cite{Kar-97} , \cite{Choubey-98}. 
But in calculating b,
as the difference in energy of two states of nuclei with the same
number of valence particles is involved those corrections are unimportant. 
The ground state energies in the $(m_p,m_n)$ spaces , given in the parentheses in Table 3,
can be compared to the experimental values but after corrections stated above are incorporated. 

The calculations also point out to a major simplification of the problem.
The changes in the values of the centroid almost fully come from the one-
body isovector part. This we have found to be true for all the cases
considered and is due to the smallness of the overall multiplicative constants
for the two-body matrix elements of both the isovector and isotensor two-body
terms given by Ormand-Brown \cite{Ormand-89}. Thus neglecting
the nonisoscalar two-body parts will be
a good approximation. Then one can easily write trace equivalent Hamiltonians
for the total Hamiltonian by adding a one-body term proportional to the
vector isospin operator ${\bf T}$ to the trace equivalent isoscalar
Hamiltonians \cite{French-69} to carry out other spectral distribution
studies.

Similar spectral distribution calculations can be carried out for the 0f-1p shell as well where 
experimental results are available \cite{Ormand-89}. There the trace subtraction procedure to project out fixed-T traces, 
as shown in equation (12), will involve many terms, particularly for low values of isospin.For this shell the interactions,
both isoscalar and nonisoscalar for which shell model studies were done \cite{Ormand-89}, are available.

\section{Conclusion}

In this paper we show that spectral distribution methods can be applied to
problems with Hamiltonians which have non-isoscalar parts also. We plan to
work out in future other examples going beyond the $1s-0d$ shell and
compare the spectral distribution results with the experimental
ones. Also replacing the isovector and
isotensor interactions by just a one-body isovector part needs detailed
future study.


\begin{table}
\begin{tabular}{|c|c|c|c|c|c|c|}
\hline
Fixed T  & Dimension & Isoscalar H & Isoscalar H & Total H  & Total H     \\
configuration  &   & centroid (MeV) & width (MeV) & centroid (MeV)& width (MeV) \\
\hline
\hline
 (400) & 105 &  -23.91  &  4.62   &  -27.31  &  4.64   \\
 (310) & 360 &  -18.07  &  4.46   &  -21.45  &  4.48   \\
 (301) & 180 &  -21.53  &  4.54   &  -24.91  &  4.56   \\
 (220) & 366 &  -12.19  &  4.50   &  -15.55  &  4.51   \\
 (211) & 408 &  -15.97  &  4.50   &  -19.33  &  4.52   \\
 (202) &  81 &  -21.07  &  4.33   &  -24.46  &  4.35   \\
 (130) & 144 &   -5.87  &  4.67   &   -9.22  &  4.69   \\
 (121) & 264 &  -10.07  &  4.66   &  -13.42  &  4.67   \\
 (112) & 120 &  -15.41  &  4.32   &  -18.77  &  4.34   \\
 (103) &  12 &  -22.40  &  3.80   &  -25.79  &  3.82   \\
 (040) &  15 &    0.23  &  5.14   &   -3.10  &  5.17   \\
 (031) &  48 &   -3.95  &  5.01   &   -7.28  &  5.03   \\
 (022) &  34 &   -9.86  &  4.57   &  -13.21  &  4.60   \\
 (013) &   8 &  -17.07  &  3.73   &  -20.42  &  3.75   \\

\hline
\end{tabular}
\caption{
Centroids and widths of the isoscalar Hamiltonian compared to the
centroids and widths of the total Hamiltonian (including the isovector
and isotensor parts) in all fixed-T configurations with m=4 particles
in $1s-0d$ shell with $T$=1. First column gives the number of particles in the particle partitions 
in the order $(0d_{5/2}^{m_1} 0d_{3/2}^{m_2}1s_{1/2}^{(m-m_1-m_2)})$.
}
\end{table}
\begin{table}
\begin{tabular}{|c|c|c|c|c|c|c|}
\hline
Fixed T & Dimension & Isoscalar H & Isoscalar H & Total H  & Total H     \\
configuration &   & centroid (MeV) & width (MeV) & centroid (MeV)& width (MeV) \\
\hline
\hline
(500) &  84 &  -31.22  &  5.30   &  -36.33  &  5.32   \\
(410) & 480 &  -25.03  &  4.91   &  -30.10  &  4.94   \\
(401) & 240 &  -28.12  &  5.33   &  -33.21  &  5.36   \\
(320) & 740 &  -19.06  &  4.87   &  -24.11  &  4.89   \\
(311) & 880 &  -22.32  &  5.11   &  -27.38  &  5.13   \\
(302) & 150 &  -27.33  &  4.90   &  -32.40  &  4.92   \\
(230) & 444 &  -12.80  &  4.99   &  -17.84  &  5.00   \\
(221) & 910 &  -16.47  &  5.18   &  -21.51  &  5.21   \\
(212) & 384 &  -21.54  &  4.78   &  -26.56  &  4.80   \\
(203) &  30 &  -27.92  &  4.12   &  -33.01  &  4.13   \\
(140) &  96 &   -6.45  &  5.29   &  -11.47  &  5.31   \\
(131) & 336 &  -10.31  &  5.45   &  -15.34  &  5.47   \\
(122) & 240 &  -15.80  &  4.94   &  -20.82  &  4.96   \\
(113) &  48 &  -22.29  &  3.92   &  -27.34  &  3.93   \\
(050) &   4 &   -0.60  &  6.23   &   -5.58  &  6.26   \\
(041) &  32 &   -4.33  &  6.00   &   -9.32  &  6.02   \\
(032) &  36 &  -10.21  &  5.29   &  -15.20  &  5.32   \\
(023) &  12 &  -17.06  &  4.21   &  -22.06  &  4.22   \\
\hline
\end{tabular}
\caption{
Centroids and widths of the isoscalar Hamiltonian compared to the
centroids and widths of the total Hamiltonian (including the isovector
and istensor parts) in all fixed-T configurations with m=5 particles
in $1s-0d$ shell with $T$=3/2.
}

\end{table}


\begin{table}
\begin{tabular}{|c|c|c|c|c|c|c|c|}
\hline
(m,T) & GSE for   & GSE for  & GSE for&$b$ from  & $b$ from &Total $|b|$ & Observed    \\
       & $H^{(0)}$&Total H & $H^{(0)}+H^{(1)}$ & GSEs  &Coulomb  & (MeV)&$|b|$      \\
      &(MeV) &(MeV) & (MeV)   &(MeV)  &(MeV) & &(MeV)      \\
\hline
\hline
(4,1) & -33.83(-33.03)   &   -37.26 (-36.44) & -37.31 &  -3.48   & -0.7   &    4.18    & 4.21  \\
(5,3/2) & -42.16 (-40.79)  &   -47.30 (-45.89) &-47.37  & -3.47   & -0.9   &    4.37    & 4.44  \\
(6,1) &  -62.88 (-61.55) &   -66.22 (-64.96)  &-66.38 &  -3.50   & -1.1   &    4.60    & 4.60  \\

\hline
\end{tabular}
\caption{
The parameter `b' of IMME coming from nuclear interactions from evaluation
of the ground state energies (GSE) calculated by
spectral distributions. The Coulomb contribution is taken from ref \cite{Lam-12}. 
The numbers in the parentheses are GSEs calculated in the $(m_p,m_n)$ spaces.
}

\end{table}

\section* {Acknowledgement}
We acknowledge Prof. Maitreyee Saha Sarkar of Saha Institute of Nuclear Physics, Kolkata, for her help during  manuscript preparation in LaTex.


\begin{thebibliography}{99}

\bibitem{Lam-12} Yi Hua Lam, N.A. Smirnova and E. Caurier, Phys. Rev. C {\bf 87}
 054304 (2013)
\bibitem{Jenkins-05} J. J\"anecke, in : Isospin in Nuclear Physics, editor D. H. Wilkinson, North-Holland Publishing, Amstetdam, 1969; D.G. Jenkins et al, Phys. Rev. C {\bf 72} 031303 (2005)

\bibitem{Wigner-57} E.P. Wigner in Proceedings of the Robert A. Welch
Foundation Conference in Chemical Research, Vol 1, edited by W.O. Milligan
(Welch Foundation, Houston, 1957) p 86

\bibitem{Ormand-89} W.E. Ormand and B.A. Brown, Nucl. Phys. A {\bf 491}
 1 (1989)
\bibitem{French-1966} J.B. French, Phys. Lett. {\bf 23}248 (1966)

\bibitem{French-67} J.B. French, Phys. Lett.B {\bf 26} 75 (1967)

\bibitem{FR-71} J.B. French,K.R. Ratcliff Phys. Rev. C {\bf 3} 94 (1971)

\bibitem{CFT-71} F.S Chang,J.B. French,and T. H. Thio, Ann.  Phys. (NY) {\bf 66} 137 (1971)

\bibitem{kothq-10} V.K.B Kota and R. U. Haq, "Spectral Distributions in Nuclei and Statistical Spectroscopy", World Scientific
Publishing Co., 2010.

\bibitem{Brody-81}
T.A. Brody, J. Flores, J.B. French, P.A. Mello, A. Pandey and S.S.M. Wong,
Rev. Mod. Phys. {\bf 53} 385 (1981)

\bibitem{French-82}
J.B. French and V.K.B. Kota, Ann. Rev. Nucl. Part. Sci. {\bf 32} 35 (1982)

\bibitem{Kota-89}
V.K.B. Kota and K. Kar, Pramana- J. Phys. {\bf 32} 647 (1989)

\bibitem{Sarkar-87} S. Sarkar, K. Kar and V.K.B. Kota, Phys. Rev. C {\bf 36} 2700 (1987)

\bibitem{Kar-97} K. Kar, S. Sarkar, J.M.G. Gomez, V.R. Mamfredi and
L. Salasnich, Phys. Rev. C {\bf 55} 1260 (1997)

\bibitem{Choubey-98} S. Choubey, K. Kar, J.M.G. Gomez and V.R. Manfredi,
Phys. Rev. C {\bf 58} 597 (1998)

\bibitem{Kar-13} K. Kar J. Phys. G {\bf 40} 015105 (2013)
\bibitem{hd-74} K.T. Hecht and J.P. Draayer, Nucl. Phys. A {\bf 223} 285 (1974)

\bibitem{French-69} J.B. French in 'Isospin in Nuclear Physics'
edited by D.H. Wilkinson (North-Holland, Amsterdam, 1969)

\bibitem{Mon-73}
K.K. Mon and J.B. French, Ann. Phys. (N. Y.) {\bf 78} 111 (1973)

\bibitem{Gomez-11}
J.M.G. Gomez, K. Kar, V.K.B. Kota, R.A. Molina, A. Relano and J. Retamosa,
Phys. Rep. {\bf 499} 103 (2011)

\bibitem{Kar-83} K. Kar, Nucl. Phys. A {\bf 368} 285 (1981)


\bibitem{Ratcliff-71} K.F. Ratcliff, Phys. Rev. C {\bf 3} 117 (1971)

\bibitem{French-66} J.B. French, in Proc. of the International School of
Physics, Enrico Fermi, Course 36, (C. Bloch ed.) Academic Press, New York (1966)


\bibitem{Mugambi-70} P.E. Mugambi, Ph.D. Thesis, University of Rochester (1970)

\bibitem{Sarkar-92} S. Sarkar, Ph.D. Thesis, University of Calcutta (1992)

\bibitem{Ormand-97} W.E. Ormand, Phys. Rev. C  {\bf 55}
 2407(1997)

\end{thebibliography}
\end{document}